\documentclass[12pt]{article}%
\usepackage{amsmath}
\usepackage{amsmath}
\usepackage{amsmath}
\usepackage{cite}
\usepackage{amsfonts}
\usepackage{amssymb}
\usepackage{graphicx}%
\csname @addtoreset\endcsname{equation}{section}
\newcommand{\eq}{\begin{equation}}
\newcommand{\feq}{\end{equation}}
\newcommand{\eqn}{\begin{eqnarray}}
\newcommand{\feqn}{\end{eqnarray}}
\newcommand{\arr}{\begin{eqnarray*}}
\newcommand{\farr}{\end{eqnarray*}}

\begin{document}
\begin{titlepage}
\vfill
\begin{flushright}
hep-th/0610047\\
\end{flushright}
\vfill
\begin{center}
\baselineskip=16pt
{\Large\bf Non-Supersymmetric Charged Domain Walls}
\vskip 1.3cm
Jan B. Gutowski$^1$  and Wafic A. Sabra $^2$
\vskip 1cm
{\small{\it
$^1$DAMTP, Centre for Mathematical Sciences\\
University of Cambridge\\
Wilberforce Road, Cambridge, CB3 0WA, UK\\}}
\vskip .6cm {\small{\it
$^2$ Centre for Advanced Mathematical Sciences and Physics Department\\
American University of Beirut\\ Lebanon  \\}}
\end{center}
\vfill
\begin{center}
\textbf{Abstract}
\end{center}
\begin{quote}
We present  general non-supersymmtric domain wall solutions with non-trivial scalar and gauge fields
for gauged five-dimensional $N=2$ supergravity
coupled to abelian vector multiplets.
\end{quote}
\vfill
\end{titlepage}

\section{Introduction}

The recent interest in the study of solutions of gauged supergravity theories
in various dimensions has to a large extent been motivated by the conjectured
Anti-de Sitter/Conformal Field Theory (AdS/CFT) equivalence \cite{maldacena}.
From the CFT perspective, supergravity vacua could correspond to an expansion
around non-zero vacuum expectation values of certain operators, or describe a
holographic renormalization group flow \cite{deboer}. \ It is hoped that this
conjectured equivalence can help in gaining some understanding of the
nonperturbative structure of gauge theories by studying classical supergravity
solutions. Of particular interest are the domain walls of gauged five
dimensional $N=2$ supergravity theories. The theory of ungauged
five-dimensional $N=2$ supergravity coupled to abelian vector supermultiplets
can be obtained by compactifying eleven-dimensional supergravity, the
low-energy limit of M-theory, on a Calabi-Yau three-fold \cite{cs}. Another
class of models are those obtained in \cite{GST} which are closely related to
Jordan algebras. The gauged five-dimensional $N=2$ supergravity theories we
consider are those obtained by gauging the $U(1)$ subgroup of the $SU(2)$
automorphism group of the superalgebra \cite{GST}. The gauging is accomplished
by introducing into the Lagrangian of the theory a linear combination of the
abelian vector fields already present in the ungauged theory, i.~e.~$A_{\mu
}=V_{I}A_{\mu}^{I}$, with a coupling constant $\chi$. The coupling of the
fermions of the theory to the $U(1)$ vector field breaks supersymmetry and
gauge-invariant terms are added to preserve $N=2$ supersymmetry. In terms of
the bosonic action of the theory, we get an additional $\chi^{2}$-dependent
scalar potential $\mathcal{V}$ \cite{GST}.

Most domain wall solutions constructed so far are configurations preserving
some of the supersymmetries (see for example \cite{some}). Explicit
supersymmetric domain wall solutions for the theories of \cite{GST}, where the
scalar fields live on symmetric spaces, were given in \cite{cks}. These
solutions, describing a holographic renormalization group flow, were expressed
in terms of Weierstrass elliptic function. Recently a systematic approach has
been employed in the classification of general supersymmetric solutions of the
gauged five dimensional supergravity with non-trivial vector multiplets
\cite{gaugedsuper, gs}. In \cite{gs}, the requirement that the scalar manifold
is a symmetric space was relaxed and the structure of solutions with null
Killing vector in both gauged and ungauged supergravity theories was also
investigated. In our present work we are interested in finding
non-supersymmetric domain wall solutions. We present non-supersymmetric
charged domain wall solutions with non-trivial scalars for all gauged
five-dimensional $N=2$ supergravity models coupled to vector multiplets. We
organize our work as follows. In section two, and in an attempt to make our
work self-contained, a brief review of the theories of the $U(1)$-gauged
supergravity and their equations of motion are given. In section three, the
domain wall solutions of \cite{cks} are presented as well as general domain
wall solutions applicable for a Calabi-Yau compactification \cite{gs}. The
analysis and the derivation of non-supersymmetric charged domain wall
solutions are given in section four. Section five includes the study of the
causal structure of the domain walls geometry and we conclude in section six.

\section{Gauged Five-Dimensional N=2 Supergravity}

The bosonic action of the gauged $D=5$ $N=2$ supergravity can be written as
\cite{GST}
\begin{equation}
S={\frac{1}{16\pi G}}\int\left(  {}R+2\chi^{2}{\mathcal{V}}-G_{IJ}F^{I}
\wedge\ast F^{J}-G_{IJ}dX^{I}\wedge\star dX^{J}-{\frac{1}{6}}C_{IJK}
F^{I}\wedge F^{J}\wedge A^{K}\right)  , \label{act}%
\end{equation}
where $I,J,K$ take values $1,\ldots,n$ and $F^{I}=dA^{I}$. Here the $X^{I}$
represent the scalar fields of the theory, which are constrained via
\[
{\frac{1}{6}}C_{IJK}X^{I}X^{J}X^{K}=1,
\]
and may be regarded as being functions of $n-1$ unconstrained scalars
$\phi^{a}$. In addition, the couplings $G_{IJ}$ depend on the scalars via
\begin{equation}
G_{IJ}={\frac{9}{2}}X_{I}X_{J}-{\frac{1}{2}}C_{IJK}X^{K}%
\end{equation}
where $X_{I}\equiv{\frac{1}{6}}C_{IJK}X^{J}X^{K}$ and therefore one has the
following useful relations
\begin{equation}
G_{IJ}X^{J}={\frac{3}{2}}X_{I}\,,\qquad G_{IJ}\partial_{a}X^{J}=-{\frac{3}{2}
}\partial_{a}X_{I}\,.
\end{equation}

The scalar potential of the gauged theory can be written in the form
\begin{equation}
{\mathcal{V}}=9V_{I}V_{J}(X^{I}X^{J}-{\frac{1}{2}}G^{IJ})
\end{equation}
where $V_{I}$ are constants \cite{GST}. The Einstein equations derived from
the action (\ref{act}) are given by
\begin{equation}
{}R_{\mu\nu}=G_{IJ}\left(  F^{I}{}_{\mu\lambda}F^{J}{}_{\nu}{}^{\lambda
}-{\frac{1}{6}}g_{\mu\nu}F^{I}{}_{\rho\sigma}F^{J\rho\sigma}+\nabla_{\mu}
X^{I}\nabla_{\nu}X^{J}\right)  -{\frac{2}{3}}\chi^{2}\mathcal{V}\,g_{\mu\nu}.
\label{e}%
\end{equation}

The Maxwell equations are
\begin{equation}
d\left(  G_{IJ}\star F^{J}\right)  +{\frac{1}{4}}C_{IJK}F^{J}\wedge F^{K}=0\,.
\label{gauge}%
\end{equation}
The scalar equations of motion give the following relations \cite{gaugedsuper,
gs}%

\begin{align}
&  d\left(  \star dX_{I}\right)  -\left(  {\frac{1}{6}}C_{MNI}-{\frac{1}{2}
}X_{I}C_{MNJ}X^{J}\right)  dX^{M}\wedge\star dX^{N}\nonumber\\
&  +\left(  X_{M}X^{P}C_{NPI}-{\frac{1}{6}}C_{MNI}-6X_{I}X_{M}X_{N}+{\frac
{1}{6}}X_{I}C_{MNJ}X^{J}\right)  F^{M}\wedge\star F^{N}\nonumber\\
&  +3\chi^{2}\left(  {\frac{1}{2}}V_{M}V_{N}G^{ML}G^{NP}C_{LPI}+X_{I}
G^{MN}V_{M}V_{N}-2X_{I}X^{M}V_{M}X^{N}V_{N}\right)  \mathrm{dvol}=0\ .
\label{fati}%
\end{align}

\section{Supersymmetric Domain Walls}

In this section we review the supersymmetric domain wall solutions found in
\cite{cks} as well as the general supersymmetric solutions with null Killing
vector and with vanishing gauge fields presented in \cite{gs}. In \cite{cks}
one starts with the following general ansatz for supersymmetric domain wall solutions:%

\begin{equation}
ds^{2}=e^{2A}(-dt^{2}+dz^{2})+e^{2B}(dr^{2}+r^{2}d\theta^{2}+r^{2}d\phi
^{2}\,),\, \label{metric}%
\end{equation}

where $A$ and $B$ are functions of the radial coordinate $r$ only. The
analysis of the Killing spinor equations (obtained from the vanishing of the
gravitini and dilatino supersymmetric variations) gives the following
restrictions on the metric and the scalar fields,%

\begin{align}
\partial_{r}A-\partial_{r}B-\frac{1}{r}  &  =0\,,\nonumber\\
\frac{1}{2}e^{-B}\partial_{r}X_{I}+\chi V_{I}+e^{-B}X_{I}(\partial_{r}%
W+\frac{1}{r})  &  =0\,. \label{scon}%
\end{align}
Eqn. (\ref{scon}) implies%

\begin{equation}
X_{I}=\frac{1}{r^{2}}e^{-2B}[-2\chi V_{I}\int e^{3B}r^{2}dr+\Lambda_{I}]\,,
\end{equation}
where the $\Lambda_{I}$ are integration constants. For symmetric spaces where
\begin{align}
C^{IJK}  &  =\delta^{II^{\prime}}\delta^{JJ^{\prime}}\delta^{KK^{\prime}
}C_{I^{\prime}J^{\prime}K^{\prime}}\,,\nonumber\\
C_{IJK}C_{J^{\prime}\left(  LM\right.  }C_{\left.  PQ\right)  K^{\prime}
}\delta^{JJ^{\prime}}\delta^{KK^{\prime}}  &  =\frac{4}{3}\delta_{I\left(
L\right.  }C_{\left.  MPQ\right)  },\nonumber\\
\mathcal{V}  &  =27C^{IJK}V_{I}V_{J}X_{K}\,,
\end{align}
it was shown that the above equations are completely integrable. Defining the
quantity \cite{cks}%

\begin{equation}
y(u)=-9a\int e^{3(B+u)}du+\frac{9}{2}\chi^{2}b\,,
\end{equation}
with%

\begin{equation}
a=C^{IJK}V_{I}V_{J}V_{K}\,,\qquad b=C^{IJK}V_{I}V_{J}\Lambda_{K}\,,\text{
\ \ \ }c=C^{IJK}V_{I}\Lambda_{J}\Lambda_{K}\,,\qquad d=C^{IJK}\Lambda
_{I}\Lambda_{J}\Lambda_{K}\,,
\end{equation}
and where the new radial coordinate $u$ is given by $u=\ln\chi r$, we obtain
the differential equation%

\begin{equation}
\left(  \frac{dy}{du}\right)  ^{2}=4y^{3}-g_{2}y-g_{3}\,, \label{weierstr}%
\end{equation}
where%

\begin{equation}
g_{2}=243\chi^{4}(b^{2}-ac)\,,\qquad g_{3}=\frac{729}{2}\chi^{6}
(3abc-a^{2}d-2b^{3})\,. \label{gg}%
\end{equation}
The general solution of Eqn.~(\ref{weierstr}) is given by $y=\wp(u+\gamma
)\,,$where $\wp(u)$ denotes the Weierstrass elliptic function, and $\gamma$ is
an integration constant.

In the classification of solutions with null Killing vector and vanishing
gauge field strengths in gauged supergravity \cite{gs}, it was found that the
metric and the scalar fields can be written in the following form
\begin{align}
ds^{2}  &  =H^{-1}\left(  2dU(dV+{\frac{1}{2}}{\mathcal{F}}dU)-(dx^{2}
)^{2}-(dx^{3})^{2}\right)  -H^{2}(dx^{1})^{2}\ ,\nonumber\\
H^{-1}X_{I}  &  =-2\chi V_{I}x^{1}+\beta_{I}(U),
\end{align}
with ${\mathcal{F}}$ \ given by
\begin{equation}
H\partial_{1}^{2}{\mathcal{F}}+H^{4}(\partial_{2}^{2}+\partial_{3}%
^{2}){\mathcal{F}}-3\partial_{1}H\partial_{1}{\mathcal{F}}={\frac{9}{2}}
H^{6}G^{IJ}\partial_{U}\beta_{I}\partial_{U}\beta_{J}\ .
\end{equation}
Hence we see that solutions for which ${\mathcal{F}}=0$ must have
$\partial_{U}\beta_{I}=0$ and hence $H$ and $X^{I}$ are also independent of
$U$.

Changing to signature $(-,+,+,+,+)$ and concentrating on solutions with
${\mathcal{F}}=0,$ we obtain the following domain wall solutions%

\begin{align}
ds^{2}  &  =H^{-1}\left(  -dt^{2}+d\omega^{2}+dx^{2}+dy^{2}\right)
+H^{2}dz^{2}\ ,\\
H^{-1}X_{I}  &  =-2\chi V_{I}z+\beta_{I}. \label{fr}%
\end{align}
Notice that these solutions are valid for all gauged $N=2$ supergravity
theories and in particular for those obtained from a Calabi-Yau
compactification. To recover the domain wall solution of \cite{cks} described
above, one can perform the following change of variable
\begin{equation}
H^{3}\left(  \frac{dz}{du}\right)  ^{2}=1,\text{ \ \ }z=\frac{b}{2\chi
a}-\frac{y}{9a\chi^{3}},
\end{equation}

\section{Non-Supersymmetric Domain Walls}

In this section we consider non-supersymmetric domain wall solutions which in
certain limits give the supersymmetric solutions considered in the previous
section. As an ansatz for non-supersymmetric solution we take:%

\begin{equation}
ds^{2}=\frac{1}{H}\left(  -fdt^{2}+dw^{2}+dx^{2}+dy^{2}\right)  +\frac{H^{2}
}{f}dz^{2} \label{nosus}%
\end{equation}
where $f$ and $H$ are functions of $z$ only. The non-vanishing components of
the Ricci tensor are given by%

\begin{align}
R_{tt}  &  =\frac{f}{2H^{5}}\left(  4fH^{\prime2}-fH^{\prime\prime}
H+f^{\prime\prime}H^{2}-4HH^{\prime}f^{\prime}\right)  ,\nonumber\\
R_{xx}  &  =R_{yy}=R_{ww}=\frac{1}{2H^{5}}\left(  -4fH^{\prime2}
+fHH^{\prime\prime}+HH^{\prime}f^{\prime}\right)  ,\nonumber\\
R_{zz}  &  =\frac{2H^{\prime\prime}}{H}-\frac{5H^{\prime^{2}}}{H^{2}}
+\frac{2H^{\prime}f^{\prime}}{fH}-\frac{f^{\prime\prime}}{2f}, \label{nonric}%
\end{align}
where the prime denotes differentiation with respect to the coordinate $z.$
The Einstein equations of motion (\ref{e}) give the following conditions:%

\begin{align}
G_{IJ}F_{zt}^{I}F_{zt}^{J}  &  =\frac{1}{2H^{3}}\left(  f^{\prime\prime}
H^{2}-3H^{\prime}f^{\prime}H\right)  ,\label{rock}\\
G_{IJ}\partial_{z}X^{I}\partial_{z}X^{J}  &  =\frac{3}{2H^{2}}\left(
H^{\prime\prime}H-2H^{\prime^{2}}\right)  ,\label{scond}\\
\chi^{2}\mathcal{V}\,  &  =\frac{3f}{4H^{4}}\left(  4H^{\prime2}%
-HH^{\prime\prime}\right)  +\frac{1}{4H^{3}}{}\left(  Hf^{\prime\prime
}-6H^{\prime}f^{\prime}\right)  \label{sf}%
\end{align}
where we allowed the gauge fields to have non-vanishing field strengths
$F_{zt}^{I}$. Note that the function $f$ drops out in (\ref{scond}) and as a
consequence we will not modify the scalars and we will simply use the ansatz
as given in the supersymmetric case (\ref{fr}). Then it can be easily
demonstrated that
\begin{align}
V_{I}X^{I}  &  =\frac{1}{2\chi}\frac{H^{\prime}}{H^{2}},\\
G^{IJ}V_{I}V_{J}  &  =\frac{1}{6\chi^{2}}\left(  \frac{H^{^{\prime\prime}}
}{H^{3}}-\frac{H^{\prime2}}{H^{4}}\right)  . \label{germ}%
\end{align}

Thus the scalar potential is given by
\begin{equation}
\mathcal{V}=\frac{3}{4\chi^{2}H^{4}}\left(  4H^{\prime2}-HH^{^{\prime\prime}
}\right)  . \label{poy}%
\end{equation}
Upon comparing (\ref{poy}) with the expression of $\mathcal{V}$ in (\ref{sf}),
the following condition is obtained%

\begin{equation}
6HH^{\prime}f^{\prime}-3\left(  1-f\right)  \left(  HH^{^{\prime\prime}
}-4H^{\prime^{2}}\right)  -f^{\prime\prime}H^{2}=0. \label{neweq}%
\end{equation}

This can be solved by%

\begin{equation}
f=1+\left(  \mu+\alpha z\right)  H^{3} \label{ns}%
\end{equation}
where $\mu$ and $\alpha$ are constants. Going back to the gauge equation of
motion (\ref{gauge}), this gives for our solution%

\begin{equation}
\partial_{z}\left(  \frac{1}{H^{2}}G_{IJ}F_{zt}^{J}\right)  =0,
\end{equation}
from which we obtain
\begin{equation}
F_{zt}^{I}=H^{2}G^{IJ}q_{J}. \label{bok}%
\end{equation}
where $q_{I}$ are constants representing electric charges. Using (\ref{rock}),
(\ref{ns}) and (\ref{bok}), we get
\begin{equation}
G^{IJ}q_{I}q_{J}=\frac{3}{2H^{4}}\left[  \alpha HH^{\prime}+\left(  \mu+\alpha
z\right)  \left(  HH^{\prime\prime}-H^{\prime2}\right)  \right]  . \label{jk}%
\end{equation}

Let us first consider the case with vanishing charges $q_{I}=0$, and take
$\alpha\neq0$. In this case, one solution of (\ref{jk}) is given by
\begin{equation}
H=\frac{c}{\left(  \mu+\alpha z\right)  } \ .
\end{equation}
Then (\ref{scond}) and (\ref{fr}) imply that the scalars are constants, with
$X_{I}=-2\frac{\chi c}{\alpha}V_{I}$.

If however, one takes $q_{I} \neq0$, with $\alpha=0$ in (\ref{jk}) then using
(\ref{germ}) we obtain the condition%

\begin{equation}
G^{IJ}\left(  q_{I}q_{J}-9\mu\chi^{2}V_{I}V_{J}\right)  =0. \label{bfc}%
\end{equation}
This can be solved by%

\begin{equation}
q_{I}=3\sqrt{\mu}\chi V_{I}. \label{ce}%
\end{equation}
Finally it remains to check whether the scalar equations of motion are
satisfied for our solution. The scalar equations (\ref{fati}) for our solution give%

\begin{align}
&  H\partial_{z}\left(  fH^{-3}\partial_{z}X_{I}\right)  -fH^{-2}\left(
{\frac{1}{6}}C_{MNI}-{\frac{1}{2}}X_{I}C_{MNJ}X^{J}\right)  \partial_{z}
X^{M}\partial_{z}X^{N}\nonumber\\
&  -H^{-1} \left(  X_{M}X^{P}C_{NPI}-{\frac{1}{6}}C_{MNI}-6X_{I}X_{M}%
X_{N}+{\frac{1}{6}}X_{I}C_{MNJ}X^{J}\right)  F^{M}_{t z} F^{N}_{t
z}\nonumber\\
&  +3\chi^{2}\left(  {\frac{1}{2}}V_{M}V_{N}G^{ML}G^{NP}C_{LPI}+X_{I}
G^{MN}V_{M}V_{N}-2X_{I}X^{M}X^{N}V_{M}V_{N}\right)  =0\ . \label{scalarnon}%
\end{align}
To simplify the calculation, we multiply the scalar equations for the
supersymmetric case, i. e. multiply
\begin{align}
&  H\partial_{z}\left(  H^{-3}\partial X_{I}\right)  -H^{-2}\left(  {\frac
{1}{6}}C_{MNI}-{\frac{1}{2}}X_{I}C_{MNJ}X^{J}\right)  \partial_{z}
X^{M}\partial_{z}X^{N}\nonumber\\
&  +3\chi^{2}\left(  {\frac{1}{2}}V_{M}V_{N}G^{ML}G^{NP}C_{LPI}+X_{I}
G^{MN}V_{M}V_{N}-2X_{I}X^{M}X^{N}V_{M}V_{N}\right)  =0\ . \label{scalarext}%
\end{align}
with $f$ and subtract the resulting equation from (\ref{scalarnon}), this
gives after using the solution for the gauge fields,
\begin{equation}
H^{\prime}\partial_{z}X_{I}+\chi^{2}\left(  4X^{K}V_{I}V_{K}-4X^{K}X^{L}
X_{I}V_{L}V_{K}\right)  H^{3}=0\ .
\end{equation}
It can be easily seen that this equation is indeed satisfied for our solution.

To summarize, we have obtained a class of \ domain wall solutions for all
gauged five-dimensional $N=2$ supergravity theories coupled to an arbitrary
number of vector multiplets. These solutions are given by%

\begin{align}
ds^{2}  &  =-\frac{\left(  1+\mu H^{3}\right)  }{H}dt^{2}+\frac{1}{H}\left(
dw^{2}+dx^{2}+dy^{2}\right)  +\frac{H^{2}}{1+\mu H^{3}}(dz)^{2}
,\label{eqn:dwsol}\\
F_{zt}^{J}  &  =3H^{2}G^{IJ}\sqrt{\mu}V_{I}\chi,\nonumber\\
X_{I}  &  =H\left(  -2\chi V_{I}z+\beta_{I}\right)  .\nonumber
\end{align}

In general, the metric is specified only implicitly by ({\ref{eqn:dwsol}}),
because $H$ is not specified explicitly by the equation of $X_{I}$. However,
when the scalar manifold is symmetric, we have the relation
\begin{equation}
\frac{9}{2}C^{IJK}X_{I}X_{J}X_{K}=1
\end{equation}
from which we can explicitly solve for $H$ and find
\begin{equation}
H=(\alpha_{0}+\alpha_{1}z+\alpha_{2}z^{2}+\alpha_{3}z^{3})^{-{\frac{1}{3}}}%
\end{equation}
where
\begin{align}
\alpha_{0}  &  ={\frac{9}{2}}C^{IJK}\beta_{I}\beta_{J}\beta_{K},\nonumber\\
\alpha_{1}  &  =-27\chi C^{IJK}V_{I}\beta_{J}\beta_{K},\nonumber\\
\alpha_{2}  &  =54\chi^{2}C^{IJK}V_{I}V_{J}\beta_{K},\nonumber\\
\alpha_{3}  &  =-36\chi^{3}C^{IJK}V_{I}V_{J}V_{K}.
\end{align}
For the special case of the $STU$ model solutions, for which the intersection
numbers are given by
\begin{equation}
C_{IJK}=|\epsilon_{IJK}|
\end{equation}
for $I,J,K=1,2,3$. In this case, $H$ factorizes as
\begin{equation}
H=(\beta_{0}z+\lambda_{0})^{-{\frac{1}{3}}}(\beta_{1}z+\lambda_{1}
)^{-{\frac{1}{3}}}(\beta_{2}z+\lambda_{2})^{-{\frac{1}{3}}}%
\end{equation}
for constants $\beta_{0},\beta_{1},\beta_{2},\lambda_{0},\lambda_{1}
,\lambda_{2}$. \ Note that as $z\rightarrow-{\frac{\lambda_{i}}{\beta_{i}}}$,
then the Ricci scalar diverges as $(\beta_{i}z+\lambda_{i})^{-{\frac{4}{3}}}$
if $\mu=0$, and as $(\beta_{i}z+\lambda_{i})^{-{\frac{7}{3}}}$ if $\mu>0$.

Hence we observe that both the supersymmetric and non-supersymmetric domain
wall solutions contain curvature singularities. However the causal structure
of the spacetimes differs considerably between the supersymmetric and
non-supersymmetric cases.

\section{Causal Structure of Domain Wall Spacetime}

To proceed, we examine the causal structure of the spacetime geometry given in
({\ref{eqn:dwsol}}) for the solutions with symmetric scalar manifolds. Observe
that geodesics on the spacetime with metric ({\ref{eqn:dwsol}}) have the
following conserved quantities%

\begin{align}
E  &  =\frac{1}{H}(1+\mu H^{3}){\dot{t}}\\
P^{i}  &  =\frac{1}{H}{\dot{x}}^{i}%
\end{align}
for $i=1,2,3$, where $t=t(\tau)$, $(x^{1},x^{2},x^{3})=(x(\tau), y(\tau),
w(\tau))$, $z=z(\tau)$, ${\dot{} }={\frac{d}{d\tau}}$ and $\tau$ is an affine
parameter. We will restrict our consideration to geodesic motion in the domain
of $z$ for which $H>0$, and we take $\mu>0$. It is convenient to define
$P^{2}=(P^{1})^{2}+(P^{2})^{2}+(P^{3})^{2}$. Then null geodesics satisfy%

\begin{equation}
\left(  {\frac{dz}{d\tau}}\right)  ^{2}=\frac{1}{H}\left(  E^{2}-(1+\mu
H^{3})P^{2}\right)  \label{nullge}%
\end{equation}
whereas timelike geodesics satisfy%

\begin{equation}
\left(  {\frac{dz}{d\tau}}\right)  ^{2}=\frac{1}{H}E^{2}-\frac{1}{H^{2}}(1+\mu
H^{3})\left(  HP^{2}+1\right)  \ . \label{timege}%
\end{equation}

Note that as $z\rightarrow\infty$, $H\sim z^{-1}$, and hence ({\ref{timege}})
implies that no timelike geodesic of fixed $E,P^{i}$ can reach $z=\pm\infty$.
In addition, causal geodesics must satisfy
\[
E^{2}-P^{2}>{\frac{E^{2}}{1+\mu H^{3}}}-P^{2}\geq{\frac{H}{1+\mu H^{3}}
}\left(  {\frac{dz}{d\tau}}\right)  ^{2}\geq0
\]
which implies that $E^{2}>P^{2}$.

\bigskip

Null geodesics of the supersymmetric solution satisfy
\begin{equation}
\left(  {\frac{dz}{d\tau}}\right)  ^{2}=\frac{1}{H}(E^{2}-P^{2}) \ .
\end{equation}
In the neighborhood of one of the curvature singularities, one can take
$H\sim\alpha z^{-{\frac{1}{3}}}$ as $z\rightarrow0$. It follows that a null
geodesic reaches the curvature singularity within finite affine parameter.
Also, null geodesics can propagate out to $z=\infty$, though they do not reach
$z=\infty$ in finite affine parameter.

Timelike geodesics of the supersymmetric solution satisfy%

\begin{equation}
\left(  {\frac{dz}{d\tau}}\right)  ^{2}=\frac{1}{H}(E^{2}-P^{2})-\frac
{1}{H^{2}} \ .
\end{equation}
Again, timelike geodesics reach the curvature singularity in finite proper
time, but are confined to lie within $0\leq z\leq z_{max}(E,P^{2})$.

Null geodesics of the non-supersymmetric solution with $P^{2}\neq0$ cannot
reach the singularity. However, null geodesics with $P^{2}=0$ satisfy
\begin{equation}
\left(  {\frac{dz}{d\tau}}\right)  ^{2}=\frac{1}{H}E^{2}%
\end{equation}
and reach the singularity in finite affine parameter. In both cases, the null
geodesics can propagate out to $z=\infty$, though they do not reach $z=\infty$
in finite affine parameter.

Timelike geodesics of the non-supersymmetric solution cannot reach the
singularity for any choice of $P$, and are therefore confined within region
$0<z_{min}(E,P^{2})\leq z\leq z_{max}(E,P^{2})$.

\section{Conclusion}

In this paper we have constructed non-supersymmetric domain wall solutions of
gauged five dimensional $N=2$ supergravity theories with non trivial vector
multiplets. The causal structure of these solutions was also discussed. These
solutions constitute generalizations to a subclass of null solutions with
vanishing gauge fields which were considered in \cite{gs}. In the
supersymmetric limit the scalar fields remain unchanged and the gauge field
strengths vanish. The scalar fields structure of these domain wall solutions
resembles those for black hole solutions considered in \cite{phase} and
therefore explicit domain wall solutions for the Calabi-Yau models considered
in \cite{phase} can be constructed.

It will be of interest to find non-supersymmetric generalizations to the
solutions of \cite{gaugedsuper, gs} and in particular to the supersymmetric
null solutions with non-trivial gauge fields of \cite{gs}. We hope to report
on this in a future publication.

\bigskip

\textbf{Acknowledgement:}

The work of W. S. was supported in part by the National Science Foundation
under grant number PHY-0313416.

\end{document}